# Real-time Detection and Tracking of Multiple Objects with Partial Decoding in H.264|AVC Bitstream Domain

Wonsang You*[a], M. S. Houari Sabirin[a], Munchurl Kim[a]

[a]Information and Communications University, Munjiro 119, 305732 Daejeon, Republic of Korea

## ABSTRACT

In this paper, we show that we can apply probabilistic spatiotemporal macroblock filtering (PSMF) and partial decoding processes to effectively detect and track multiple objects in real time in H.264|AVC bitstreams with stationary background. Our contribution is that our method cannot only show fast processing time but also handle multiple moving objects that are articulated, changing in size or internally have monotonous color, even though they contain a chaotic set of non-homogeneous motion vectors inside. In addition, our partial decoding process for H.264|AVC bitstreams enables to improve the accuracy of object trajectories and overcome long occlusion by using extracted color information.

## 1. INTRODUCTION

Most of object detection and tracking techniques, called the *pixel domain algorithms*, fundamentally utilize raw pixel data to guarantee the accuracy of object trajectories. However, they tend to cause high computational complexity which is fatal to large-scale surveillance systems where multiple cameras operate synchronistically at a fast pace or in real time under restricted hardware circumstance. Moreover, most of video contents transmitted over networks are not original pixel data but a sort of bitstreams encoded by a compression tool like MPEG to enhance the efficiency of communication. In that case, the pixel domain algorithms require additional computation to fully decode such bitstreams.

On the contrary, the *compressed domain algorithms* can significantly reduce computational complexity by exploiting encoded information such as motion vector and DCT coefficients. Nevertheless, their performance of object detection and tracking has been considerably worse than pixel domain algorithms since they make use of only the encoded information which is not reliable enough to be employed as a clue for object detection and tracking. Such a problem originates from the fact that motion vectors extracted from a bitstream are sparsely assigned to blocks instead of pixels, and do not exactly correspond to real motion of objects. In that reason, these algorithms are usually applicable just to restricted situations such that motion vectors inside each object are sufficiently reliable and relatively homogeneous. Moreover, they have not efficiently handled occlusion problem since it is hard to distinguish several objects just by using encoded information. To compensate the weaknesses, some compressed domain algorithms utilize a low-resolution image which can be easily extracted from DCT coefficients without full decoding [2,5,6,8-10,14]. Unfortunately, it is impossible for MPEG-4 AVC compressed videos [1], which makes it difficult to apply any AVC-based algorithm to surveillance systems where several objects are required to be identified on the basis of color information.

To address these problems especially in H.264|AVC bitstreams with more natural scenes than those handled by previous compressed domain algorithms, we propose a hybrid algorithm of compressed domain and pixel domain which rapidly detects and tracks an a priori unknown number of moving objects with stationary background. As an alternative to low-resolution images, we utilize partially-decoded pixel data of object blob regions. It involves two hierarchical phases:

1. *Extraction Phase*: We roughly extract the block-level region of objects and construct the approximate object trajectories in each P-frame by the PSMF. We first eliminate probable background macroblocks, and then cluster the remaining macroblocks into several fragments. To distinguish object fragments from background fragments, we use two steps of filtering. At the first step, we filter off background fragments on the basis of integer transform (IT) coefficients and spatial structure of fragments. At the second step, we observe temporal consistency of each surviving fragment over a given period, and approximate the probability that each fragment would be a part of objects. The fragments with high occurrence probability are finally considered as a part of objects.

*wonsang.you@informatik.uni-augsburg.de; phone 49 821 598 2378; fax 49 821 598 4329; multimedia-computing.org

2. *Refinement Phase*: We then accurately refine the object trajectories which are roughly generated in the extraction stage. First, we partially decode only object blobs in each I-frame, and perform background subtraction in each I-frame and motion interpolation in each P-frame. Then, the color information of each object is extracted and recorded in a database so that we can identify each object in spite of its long occlusion or temporary vanishment.

In contrast to most compressed domain algorithms, the proposed hybrid algorithm guarantees reliable performance especially in such complicated scenes that multiple objects are fully occluded each other for a long time or that an object has flat color distribution or internally non-rigid motion. Moreover, we first introduces an approximate partial decoding method in I-frames for H.264|AVC bitstreams with standstill background, which has been deemed to be primarily impossible due to spatial prediction dependency on neighboring blocks.

## 2. RELATED WORK

The conventional compressed domain methods can be divided into clustering-based methods and filtering-based methods. While the former emphasizes the local similarity of blocks, the latter puts emphasis on the global similarity in a whole image.

### 2.1 Clustering-based Approaches

The clustering-based methods first attempt to split a frame into several block-level fragments on the basis of the homogeneity of motion vectors or DCT coefficients [3,6,8,9,11-13]. Then, each fragment is merged into a similar neighboring fragment, and is finally classified as background or foreground.

The most preferential clue for block clustering has been the similarity of motion vectors [10,11]. However, since motion vectors do not always correspond to optical flow, such a block clustering is not credible. In [9], the unreliability of motion vectors is settled by filtering off unreliable motion vectors. Motion vectors have another drawback that the motion vector field is too sparse to cause inaccurate object segmentation and tracking; it can be solved by spatial interpolation and expectation maximization (EM) as shown in [13]. Some algorithms mainly exploit a low resolution image, called a *DC image*, which are constructed out of DC DCT coefficients in I-frames, rather than motion vectors [2,5,6,8-10,14]. Note that it is impossible to make DC images in I-frames of a H.264|AVC bitstream due to spatial prediction dependency on neighboring blocks [1]. The most advanced clustering-based method is the region growing approach that several seed fragments grow spatially and temporally by merging similar neighboring fragments [18,19].

### 2.2 Filtering-based Approaches

The filtering-based methods first extract the foreground region by removing such blocks that are unreliable or judged to belong to background. Once the global segmentation is completed, the foreground region is split into multiple objects.

Spatiotemporal confidence measurement of motion vectors and DCT coefficients can be employed to filter off unreliable blocks [7]. Global motion compensation and background subtraction based on DC images are also beneficial to extract the foreground region [2,4,5,20]. Recently, the advanced algorithms, based on the Markovian random field (MRF) theory and the Bayesian estimation framework, have been proposed. They maximize the predefined probability to find out the optimal configuration of block-level object regions [6,12,14,15]. In general, these algorithms have fairly reliable performance, but they also have high computational complexity, which detracts from the merit of the compressed domain approach. Currently, there are three algorithms which handle H.264|AVC compressed videos: the MRF-based algorithm [12], the dissimilarity minimization (DM) algorithm [16], and the probabilistic data association filtering (PDAF) algorithm [14].

Although our method has many features in common with the filtering-based approach, it is exactly different from them in three respects. First, we employ the PSMF, which is less complicated than the MAF-based and PDAF-based algorithms, to significantly reduce the computational complexity. Second, our method exploits partially-decoded pixel data in I-frames as well as encoded information to effectively complement the PSMF which may lead to inaccurate object trajectories. The DM algorithm also makes use of partially-decoded pixel data; however, it may not have a significant effect on reducing computational complexity since it requires not merely partial decoding in all P-frames but also full decoding in each I-frame. Furthermore, it does not support the automatic detection of objects. Third, our method is not premised on the homogeneity of motion vectors inside an object; it allows us to apply our method to manifold situations such as an object which is articulated or changing in size.

# 3. PROBABILISTIC SPATIOTEMPORAL MACROBLOCK FILTERING

The probabilistic spatiotemporal macroblock filtering (PDMF) is the process of filtering background macroblocks on the basis of their spatial and temporal properties, in order to rapidly segment object regions in the macroblock unit and track each object roughly in real time. The process is organized as blocks clustering, spatial and temporal filtering.

## 3.1 Block Clustering

We assume that a video is recorded by a fixed camera under an environment devoid of illumination change, and encoded with the AVC baseline profile especially where one I-frame is periodically inserted less than every 10 frames. Also, each visible object is supposed to separately appear in a scene at the start and moves nearly at a foot's pace, and to occupy at least more than two macroblocks. In that case, it is observed in P-frames that most parts of the background tend to be encoded into skip macroblocks since they theoretically have no residual error between the motion-compensated prediction block and the source macroblock. Contrary to background macroblocks, most parts of objects which prominently move in a scene tend to be encoded into non-skip macroblocks since most of macroblocks inside the moving objects are split into several sub-blocks or have residual errors in the motion prediction of encoding process due to dynamic change in shape or color. Thus, we can reduce search space by filtering off all skip macroblocks which are considered as potential parts of the background. Then, the remaining macroblocks are naturally clustered by their mutual connectivity as depicted in Fig. 1. In other words, we obtain several fragments, called *block groups*, which consist of non-skip macroblocks connected in the horizontal, vertical, or diagonal directions. The process is called *block clustering*.

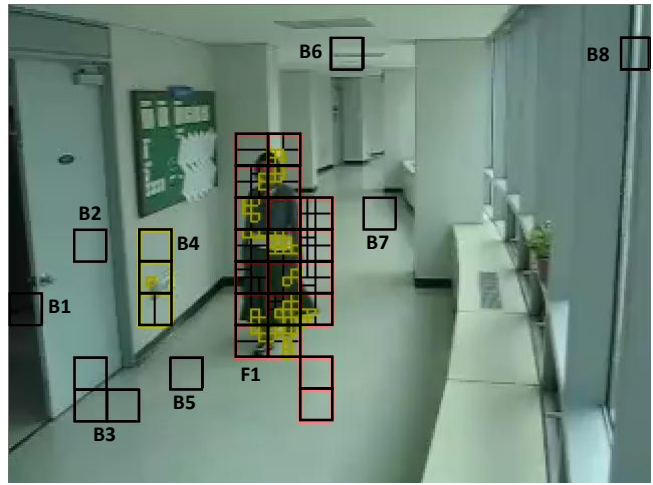

Fig. 1. Block clustering and spatial filtering. Each black rectangle box represents a non-skip macroblock which consists of one or several sub-blocks while white small rectangle boxes represent 4x4 unit blocks with non-zero IT coefficients.

In terms of block groups, we can formulate our subsequent task as twofold. The first task is to further eliminate erroneous block groups which substantially belong to the background, and to merge homogeneous block groups of the foreground into one independent object. The second task is to correct the miscalculation of object trajectories which can occur by accidentally filtering off even skip blocks inside the foreground. We use spatial filtering and temporal filtering to work out the first task, and use background subtraction and motion interpolation for the second task which is discussed in more details in the section 4.

## 3.2 Spatial Filtering

We have observed that most block groups, which consist of just one isolated macroblock or do not contain any non-zero IT coefficient, belong to the background. In this reason, we additionally filter off such block groups; the process is called *spatial filtering* and the surviving block groups are called *active block groups*. It should be noticed that there is a trade-off between performance and computational complexity, which may be controlled according to whether to carry out spatial filtering. If we skip the spatial filtering, we can detect even extremely uncommon objects which occupy less than a macroblock or do not have non-zero IT coefficients. However, it brings about an appreciable increase in computation complexity of temporal filtering while it negligibly improves the performance of object detection.

The block clustering and spatial filtering is illustrated in Fig. 1. Nine block groups (indicated as B1~B8, and F1) emerge from a frame as a result of block clustering. Once spatial filtering is applied to them, only two active block groups (F1, B4) are left. Most block groups (B1, B2, B5~B8) are removed since they are composed of just one isolated macroblock while a block group (B3) is eliminated due to its all zero IT coefficients.

### 3.3 Temporal Filtering

To get rid of erroneous active block groups (such as B4 in Fig. 1) which remain after spatial filtering, we supplementally execute *temporal filtering* which rejects such active block groups that are not consistent in their appearance and have low occurrence probability over a given frame interval called the *observation period*. In particular, some active block groups, whose location and motion has coherent spatiotemporal trends, are merged into one single object. The structure of temporal filtering is illustrated in Fig. 2. Unless an active block group is not overlapped with any active block group in the preceding frame, it is considered to be newly detected. In that case, it is assigned to a new *entity l* with the initial label of *candidate object* **C**, and is named the *seed region* of the entity. To trace each entity, we project the seed region onto the subsequent P-frame, and then search a set of *active block groups*, which are overlapped with the projected region, called the *succeeding region*. In this manner, we recursively compute the succeeding regions of the entity in later P-frames. Given the seed region $\mathbf{G}_l^1$, the succeeding region of the entity *l* in the *i*th frame ($i > 1$) is defined as follows:

$$\mathbf{G}_l^i = \left\{ \mathbf{X} \middle| \mathbf{X} \cap \hat{\mathbf{G}}_l^{i-1} \neq \phi, \mathbf{X} \subset \mathbf{C}^i \right\} \quad (1)$$

where $\hat{\mathbf{G}}_l^{i-1} = \mathbf{G}_l^{i-1} \cup \overline{\mathbf{G}}_l^{i-1}$, $\mathbf{C}^i$ denotes the set of all active block groups in the *i*th frame, and **X** is an active block group. In the case that there are no active block groups which are overlapped with the succeeding region in the preceding frame, that is, if $\mathbf{G}_l^i = \phi$, we set a *virtual active block group* $\overline{\mathbf{G}}_l^i$ where $\overline{\mathbf{G}}_l^i = \mathbf{G}_l^{i-1}$ as depicted in $\overline{G}_6^3$ of Fig. 2. Otherwise, $\overline{\mathbf{G}}_l^i$ is set to be an empty set.

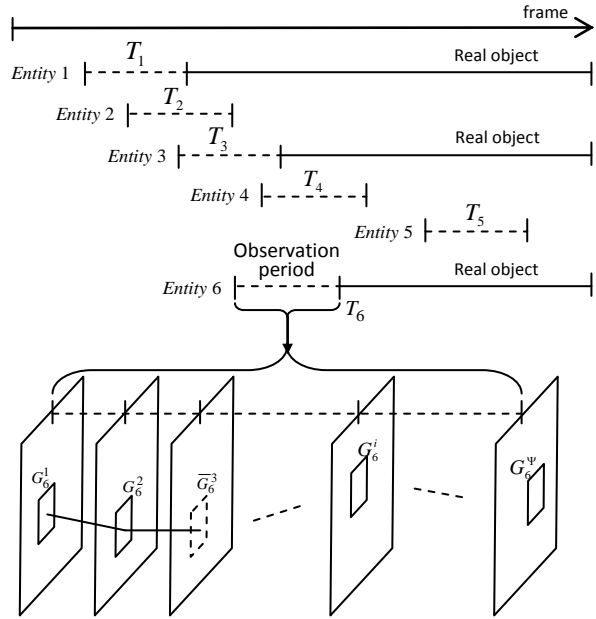

Fig. 2. Temporal filtering. Each bar shows how long an entity lasts over frames. Once a new active block group (*Entity* 1~6) is found in a frame, the corresponding active group train ($T_1 \sim T_6$) is built up during the observation period as indicated by a wavy line. A small box inside each frame, shown at the bottom, represents an active block group as a component of the active group train. Then, three bars of entities (*Entity* 1, 2, and 6) considered as a real object are expanded.

Thereafter, to judge whether each entity represents an object or the background according to its temporal coherence for the observation period, we define the *active group train* for the entity *l* as follows:

$$\mathbf{T}_l = \{\mathbf{X} | \mathbf{X} = \hat{\mathbf{G}}_l^i, i = 1, \ldots, \Psi\} \tag{2}$$

where $\Psi$ indicates the length of the observation period. Then, we calculate the occurrence probability $\mathbf{P}_l$ that the entity $l$ belongs to an object given the active group train. If the occurrence probability is higher than a predefined threshold, the entity $l$ is regarded as a *real object*. Otherwise, it is considered to be a part of the background and then is eliminated. In this manner, we have the following criterion:

$$\mathbf{P}_l = \mathbf{P}(l = \mathbf{R}|\mathbf{T}_l) > \Omega \tag{3}$$

where R is the label of real object and $\Omega$ is the threshold. According to the Bayes rule, we have

$$\mathbf{P}(l = \mathbf{R}|\mathbf{T}_l) = \mathbf{P}(l = \mathbf{R}|\hat{\mathbf{G}}_l^1, \hat{\mathbf{G}}_l^2, \ldots, \hat{\mathbf{G}}_l^\Psi) = \frac{\mathbf{P}(l = \mathbf{R}, \hat{\mathbf{G}}_l^1, \hat{\mathbf{G}}_l^2, \ldots, \hat{\mathbf{G}}_l^\Psi)}{\mathbf{P}(\hat{\mathbf{G}}_l^1, \hat{\mathbf{G}}_l^2, \ldots, \hat{\mathbf{G}}_l^\Psi)}$$
$$= \prod_{i=1}^{\Psi} \mathbf{P}(\hat{\mathbf{G}}_l^i | \hat{\mathbf{G}}_l^{i-1}, \ldots, \hat{\mathbf{G}}_l^1, l = \mathbf{R}) \frac{\mathbf{P}(l = \mathbf{R})}{\mathbf{P}(\hat{\mathbf{G}}_l^1, \hat{\mathbf{G}}_l^2, \ldots, \hat{\mathbf{G}}_l^\Psi)} \tag{4}$$

In (4), $\mathbf{P}(l = \mathbf{R})$ and $\mathbf{P}(\hat{\mathbf{G}}_l^1, \hat{\mathbf{G}}_l^2, \ldots, \hat{\mathbf{G}}_l^\Psi)$ are invariable as *a priori* probabilities. Thus, we have

$$\mathbf{P}(l = \mathbf{R}|\mathbf{T}_l) \propto \prod_{i=1}^{\Psi} \mathbf{P}(\mathbf{G}_l^i | \hat{\mathbf{G}}_l^{i-1}, \ldots, \hat{\mathbf{G}}_l^1, l = \mathbf{R}) \tag{5}$$

Accordingly, our criterion of (3) is simplified as follows:

$$-\sum_{i=1}^{\Psi} \ln \mathbf{P}(\hat{\mathbf{G}}_l^i | \hat{\mathbf{G}}_l^{i-1}, \ldots, \hat{\mathbf{G}}_l^1, l = \mathbf{R}) < \omega \tag{6}$$

where $\omega$ is called the *occurrence threshold* such that $\omega > 0$. In the case of $\mathbf{G}_l^i \neq \phi$, we suppose that the succeeding region of the entity $l$ depends on the history just in the preceding frame. Then, $\mathbf{P}(\hat{\mathbf{G}}_l^i | \hat{\mathbf{G}}_l^{i-1}, \ldots, \hat{\mathbf{G}}_l^1, l = \mathbf{R})$ is inferred as follows:

$$\mathbf{P}(\hat{\mathbf{G}}_l^i | \hat{\mathbf{G}}_l^{i-1}, \ldots, \hat{\mathbf{G}}_l^1, l = \mathbf{R})$$
$$= \mathbf{P}(\hat{\mathbf{G}}_l^i | \hat{\mathbf{G}}_l^{i-1}, l = \mathbf{R}) = \mathbf{P}(\mathbf{G}_l^i | \hat{\mathbf{G}}_l^{i-1}, l = \mathbf{R}) = \frac{n(\mathbf{G}_l^i \cap \hat{\mathbf{G}}_l^{i-1})}{n(\hat{\mathbf{G}}_l^{i-1})} \tag{7}$$

where $n(\mathbf{G}_l^i)$ denotes the number of macroblocks in $\mathbf{G}_l^i$, and $n(\mathbf{G}_l^i \cap \hat{\mathbf{G}}_l^{i-1})$ is that of macroblocks in $\mathbf{G}_l^i$ which are overlapped with $\hat{\mathbf{G}}_l^{i-1}$. If $\mathbf{G}_l^i = \phi$, $\mathbf{P}(\hat{\mathbf{G}}_l^i | \hat{\mathbf{G}}_l^{i-1}, \ldots, \hat{\mathbf{G}}_l^1, l = \mathbf{R})$ is estimated as follows:

$$\mathbf{P}(\hat{\mathbf{G}}_l^i | \hat{\mathbf{G}}_l^{i-1}, \ldots, \hat{\mathbf{G}}_l^1, l = \mathbf{R}) = \frac{o(\hat{\mathbf{G}}_l^i, \hat{\mathbf{G}}_l^{i-1}, \ldots, \hat{\mathbf{G}}_l^1)}{i} \tag{8}$$

where $o(\hat{\mathbf{G}}_l^i, \hat{\mathbf{G}}_l^{i-1}, \ldots, \hat{\mathbf{G}}_l^1)$ is the number of frames in which the succeeding region of the entity $l$ is detected within the observation period. It is based on the assumption that any object continues to move visually without an abrupt halt or vanishment during the observation period since it is newly detected. Hence, if the active block groups corresponding to a candidate object rarely appear during the observation period, the candidate object is considered as the background.

Once an entity is regarded as a real object by the above criterion (6), we continue to track it frame by frame after the observation period in the same manner as (1). Although the active block groups corresponding to any real object are not found for a long time, the real object is assumed to remain motionless without its disappearance; thus, its succeeding region exists as a type of virtual active block group.

## 4. REFINEMENT AND RECOVERY OF OBJECT TRAJECTORIES

Suppose that the blob of a real object in the $i$th frame is represented by the *feature vector* $\mathbf{F}^i = (\mathbf{p}^i, h^i, w^i)$ where $\mathbf{p}^i$ denotes the location of the real object and $(h^i, w^i)$ is the size of the real object with its height and width. The feature vector can be primarily determined by a rectangle that encompasses the exterior of all active block groups which compose the real object. However, the result is not precise since some parts of objects can be excluded from the segmented foreground region in the block clustering process as discussed in the section 3.1. To accurately compute the trajectories of objects, we employ background subtraction in I-frames and motion interpolation in P-frames. In other words, once we get more accurate location and size of each object in an I-frame by background subtraction, we go back to the previous P-frames between two subsequent I-frames and correctly revise the blob information in each P-frame by motion interpolation. It allows us to periodically update the trajectories every Group of Pictures (GOP).

### 4.1 Partial Decoding and Background Subtraction

The background subtraction process is organized into three steps as shown in Fig. 3. First, we approximately predict the location and size of an object blob in an I-frame. Second, we decode only the estimated region of object blobs (Fig. 3(c)) instead of the entire region (Fig. 3(b)) in an I-frame. Third, the partially decoded region is subtracted from the background image (Fig. (a)), and then the final blob is decided to be the rectangle which most tightly encompasses the segmented region as shown in Fig. 3(d).

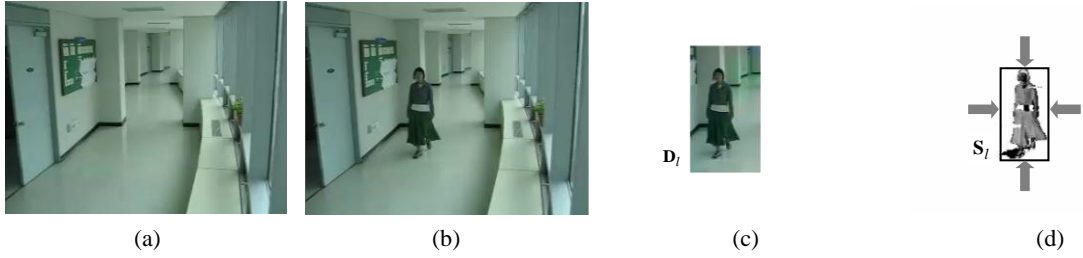

(a) (b) (c) (d)

Fig. 3. Background subtraction in an I-frame. (a) Background image. (b) Original image. (c) Partially-decoded region $D_l$. (d) The segmented object region $S_l$ after background subtraction is applied to the original image. The rectangle box, fitted into the segmented region, is finally decided to be the object blob corresponding to the current I-frame.

In the first step, the initial location and size of an object blob in an I-frame is predicted on the basis of blob information in P-frames of the previous GOP as follows:

$$\mathbf{F}^i = \left( \mathbf{p}^{i-1}, \max_{1 \le k \le N-1} h^{i-k}, \max_{1 \le k \le N-1} w^{i-k} \right) \qquad (9)$$

where N denotes the length of a GOP. The location is assumed to be the same as the location in the previous P-frame since the visible change of an object, which moves at a foot's pace and is remote from the camera, is generally negligible between two subsequent frames in normal frame rate (30 frames/second). The height and width are predicted to be the respective maximum of height and width in P-frames of the previous GOP. It allows us to improve the likelihood that the predicted blob encompasses the entire region of an object.

In the second step, partial decoding enables us to significantly reduce the computational complexity for decoding process. Note that it is effective when the foreground occupies less than a half of the entire picture. Originally, partial decoding in I-frames of a H.264|AVC bitstream has been known to be impossible since decoding a unit block (16x16 macroblock or 4x4 sub-macroblock) in an I-frame requires its spatial prediction which depends on pixels of its neighboring blocks [1]. In other words, to decode a block in an I-frame, its neighboring blocks should also be decoded *a priori*. In the worst case that the most bottom-right block is decoded, a lot of blocks which are located leftward or upward have to be decoded *a priori*, which leads to an increase in computational complexity. To avoid such a problem, we substitute the reference pixels of the neighboring blocks with the pixels of the initial background image. It is restrictively applicable to the scene environment which has a stationary background and no illumination change.

In the last step, since our background model is uncomplicated, we employ the basic background subtraction which is performed by thresholding the absolute difference between the partially decoded image and the background image as follows:

$$\mathbf{S}_l = \{\mathbf{x} \mid |\mathbf{I}(\mathbf{x}) - \mathbf{B}(\mathbf{x})| > \varepsilon, \mathbf{x} \in \mathbf{D}_l\} \quad (10)$$

where $\mathbf{S}_l$ is the segmented region of the object $l$ as depicted in Fig. 3(d), $\mathbf{I}(\mathbf{x})$ is the pixel of the location $\mathbf{x}$ in the partially decoded image, $\mathbf{B}(\mathbf{x})$ is the pixel of $\mathbf{x}$ in the background image, $\varepsilon$ is the threshold, and $\mathbf{D}_l$ denotes the partially decoded region of the object $l$ as depicted in Fig. 3(c). The noise in $\mathbf{S}_l$ can be effectively cleared by connected component analysis and morphological filtering [21]. Then, the blob of the object $l$ is finally determined to be a rectangle which tightens the segmented region.

### 4.2 Motion Interpolation

As discussed in the section 3.1, the PSMF-based object tracking method can cause the prominent discrepancy between the actual object region and the estimated macroblock-level region which consists of active block groups. The size and location of each object blob can vary factitiously over P-frames as shown in the dotted rectangle boxes in Fig. 4.

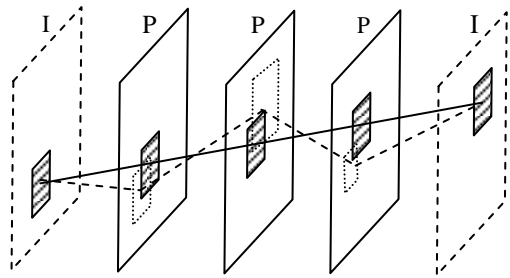

Fig. 4. Motion interpolation in P-frames. The dotted rectangle boxes indicate roughly-estimated object blobs while the shaded rectangle boxes represent accurate object blobs which are corrected through motion interpolation.

To compensate this defect of the PSMF, we introduce the motion interpolation method in P-frames in addition to background subtraction described in the former section. Under our assumption that the length of a GOP is less than ten frames and each object moves slowly at a foot's pace, it is clear that the linearly interpolated feature vector for each blob is more accurate than the rough feature vector estimated by the PSMF as illustrated in the shaded rectangle boxes in Fig. 4. It can be computed as follows:

$$\mathbf{F}^{i-k} = \mathbf{F}^i + \frac{k}{N}\left(\mathbf{F}^{i-N} - \mathbf{F}^i\right) \quad (11)$$

where $k$ ($0 < k < N$) is the index for P-frames. It should be noticed that as the length of one GOP gets longer, the updated feature vectors are less reliable because the linearity assumption no longer holds true.

### 4.3 Occlusion Handling

We introduce here a method for perceiving occlusion and disocclusion and identifying each object. The *occlusion*, in which two or more independent objects are occluded each other, can be modeled as the *region collision* of active block groups as shown in Fig. 5(a). The region collision occurs under the situation that an active block group is overlapped with more than one active block group in the preceding frame. Let us assume that an active block group $\Lambda^i$ in the $i$th frame is overlapped with a set of active block groups $\mathbf{M}^{i-1} = \{\mathbf{G}_1^{i-1}, \ldots, \mathbf{G}_n^{i-1}\}$ in the preceding frame. If $\mathbf{M}^{i-1}$ includes only one real object, $\Lambda^i$ is regarded as the identical entity with it and becomes the succeeding region of the real object. On the contrary, when $\mathbf{M}^{i-1}$ includes two or more real objects, the region collision can be regarded as occlusion. In that case, we suppose that the entity of $\Lambda^i$ is not a real or candidate object but in the state of occlusion; however, it is tracked in the same way as real objects until the occlusion terminates. In either case, all candidate objects which belong to $\mathbf{M}^{i-1}$ are merged to other subsistent real objects. Note that once the occlusion occurs, we store up the hue color histogram of each object region, which is subtracted from the background in the last I-frame just before occlusion, called the *prior color distribution*, so that we can identify each object after disocclusion. The hue color component is a good parameter for distinguishing different colors especially under surveillance environments because it is invariant to illumination change [17].

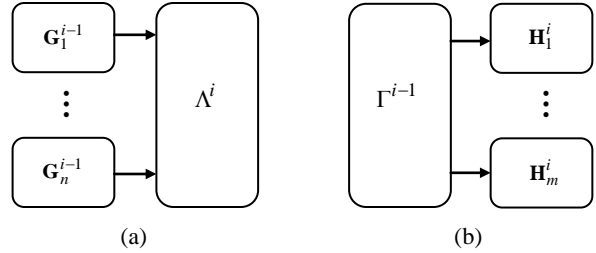

Fig. 5. The model of occlusion and disocclusion. (a) Region collision. (b) Region split. Each box represents an active block group.

On the other hand, the *disocclusion* can be modeled as the *region split* of one active block group as shown in Fig. 5(b). The region split occurs under the situation that two or more active block groups are simultaneously overlapped with an occlusion entity in the preceding frame. In the case that a set of active block groups $\mathbf{N}^i = \{\mathbf{H}_1^i,...,\mathbf{H}_m^i\}$ are overlapped with one occlusion entity $\Gamma^{i-1}$ in the preceding frame, we can expect two feasible cases. First, it can be just a transient separation of the occlusion region which is shortly followed by reunion. Second, each element in $\mathbf{N}^i$ can be regarded as either an erroneous background partition or a part of previously or newly detected real objects. To examine which scenario correspond to the current state of region split, we apply temporal filtering to them by the same way as normal candidate objects discussed in the former section. If at least two elements in $\mathbf{N}$ turn out to be among real objects after the observation period, the region split can be regarded as disocclusion. To identify each disoccluded real object, we store up the hue color histogram of each object region in the first I-frame just after disocclusion, called the *posterior color distribution*. Then, we use the Euclidean distance of two hue color histograms to compare the posterior color distribution of each real object with the prior color distributions of real objects which are previously detected before occlusion. The real object, whose color distribution has the shortest Euclidean distance from that of the target object, is considered as the best matched one.

## 5.  EXPERIMENTAL RESULTS

### 5.1 Video Sequences

Two video sequences, which were taken by one fixed camera indoors and outdoors, were used to test the proposed algorithms. While only one person walking in a corridor of a university building appears in the indoor sequence, three persons entering individually into the visual field of the camera appears in the outdoor sequence without visual occlusion. In two sequences, there was no illumination change of the background. Each sequence was encoded with 320x240 size at 30 frames per second by the JM12.4 reference software with the GOP structure of 'IP···PIP···' based on the AVC baseline profile. Especially, all P-frames were set to have no intra-coded macroblocks. Also, the length of observation period was set to be 8 frames.

### 5.2 Extraction Phase

In the spatial filtering process, 70.8% of block groups in the indoor sequence and 64.2% in the outdoor sequence were filtered off on average. The block groups, which contain just one isolated macroblock or do not have non-zero IT coefficients, were successfully eliminated by spatial filtering. Fig. 6 illustrates the result of temporal filtering in the outdoor sequence. Only three entities survived among 27 active group trains and were labeled 'real object', which coincides with the actual situation. On the contrary, other entities were labeled 'background' or merged into neighboring real objects. As a result of temporal filtering, 96% of all active group trains were removed in the whole sequence.

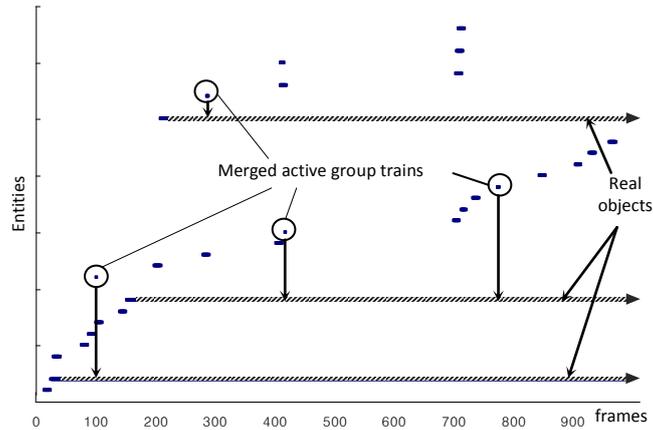

Fig. 6. The illustration of temporal filtering. Short bars indicate active group trains whose entities turned out to be a part of the background or get merged into their neighboring real object. Three long bars imply that three entities were decided as a real object.

### 5.3 Refinement Phase

To obtain more precise object trajectories, we have used background subtraction in I-frames and motion interpolation in P-frames as explained in the section 4. Fig. 7 shows the results in three steps of background subtraction in I-frames of the indoor and outdoor sequences: partial decoding, foreground extraction, and blob optimization. The background subtraction in our experiments did not involve morphological filtering for noise removal. Fig. 8 illustrates how accurately the motion interpolation corrects the location and size of an object blob in a P-frame. In the PSMF-based extraction phase before motion interpolation, the rectangle box of the blob did not enclose some parts of the object as shown in Fig. 8(a). It was well refined by motion interpolation as shown in Fig. 8(b).

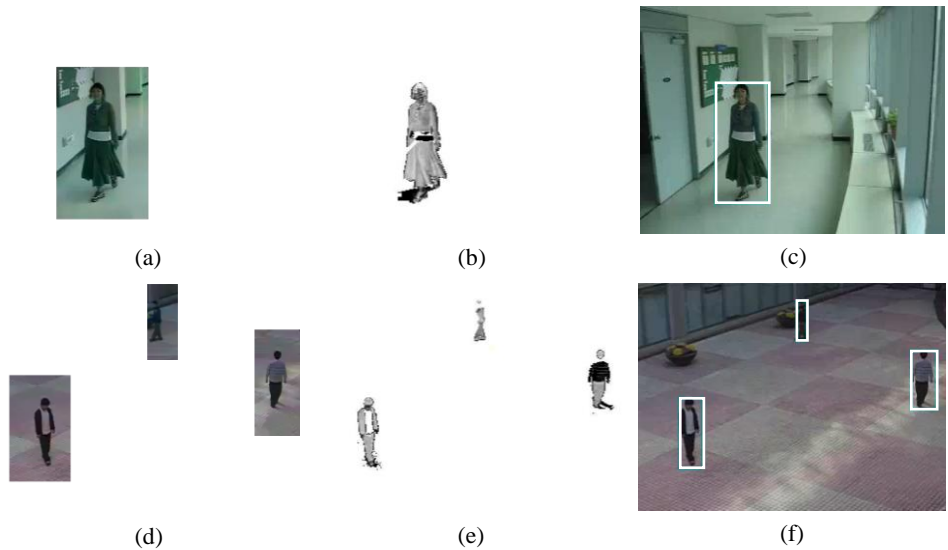

Fig. 7. The result of background subtraction in I-frames of both indoor and outdoor sequences. (a) and (c) shows partially-decoded images in I-frames; on the other hand, (b) and (e) shows their background-subtracted images. Lastly, (c) and (f) depict the object blobs optimized by background subtraction.

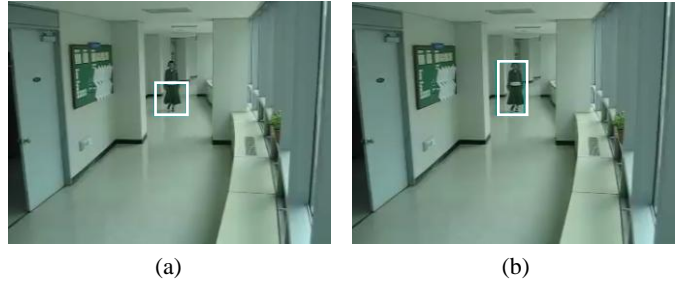

(a)　　　　　　　　　　　　　(b)

Fig. 8. The effect of motion interpolation in a P-frame of the Indoor Sequence. (a) An inaccurate object blob generated by the PSMF. (b) The object blob revised after motion interpolation.

### 5.4 Performance Analysis and Occlusion Handling

The proposed method exhibited a satisfactory performance over 720 and 990 frames of the indoor and outdoor sequences, respectively. It can be noticed in Fig. 9(a) that the performance in the indoor sequence was kept good even though the object was continually changing in size as a person was moving toward the camera-looking direction. Moreover, although the body parts (such as head, arms, and legs) of the person had different motion, the rectangle box of the object blob always enclosed the whole body precisely. Likewise, even in the outdoor sequence which contains multiple objects as shown in Fig. 9(b), the proposed algorithm did not fail to detect and track three persons separately.

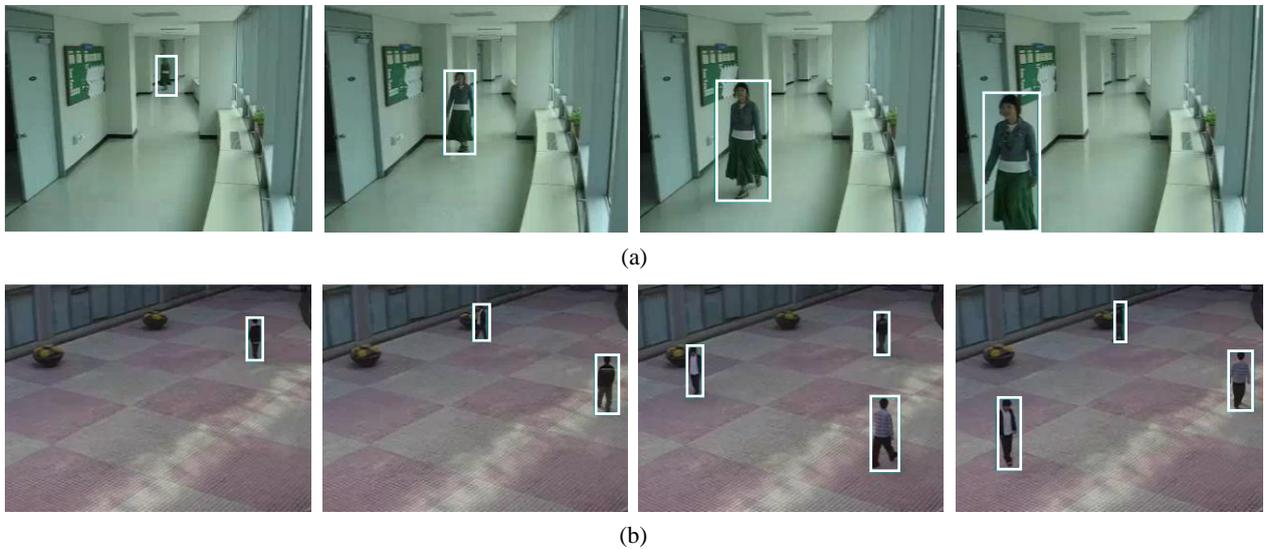

(a)

(b)

Fig. 9. The result of object detection and tracking. (a) The indoor sequence such that one person walks down a hallway. (b) Outdoor sequence such that three persons walks down an open space simultaneously.

In the case of object occlusion, we used the hue color histograms, whose components were quantized into 64 bins, to identify each object after disocclusion as discussed in the section 4.3. In other words, the hue color histogram of a disoccluded object was compared with that of each object which was already detected before occlusion on the basis of the Euclidean distance of two hue color histograms. Then, the object, whose color distribution had the smallest Euclidean distance, was considered to be the same as the disoccluded object. As shown in Fig. 10, two objects were very well distinguished by the normalized hue color histograms.

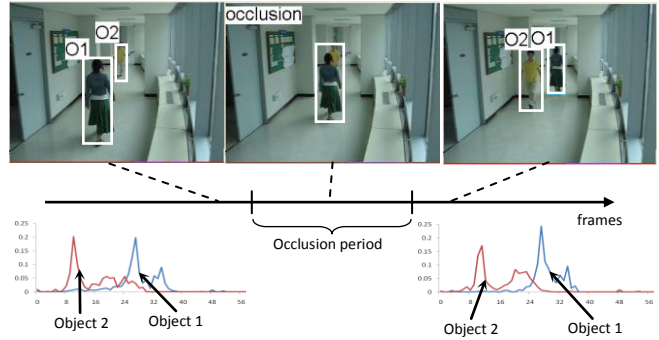

Fig. 10. The result of occlusion handling. Three images show the object blob before occlusion, the object blob in the middle of occlusion, and the object blob after disocclusion, respectively. The above graphs represent each person' hue color histograms before occlusion and after disocclusion.

### 5.5 Computational Cost

The computation of the proposed algorithm involves three major processes: partial decoding in I-frames, the extraction of macroblock types and IT coefficients in P-frames, and object detection and tracking. Especially, the computation time in the first process is greatly influenced by the type of AVC decoder. As a result, the processing time were taken 2.02 milliseconds per frame (49.5 frames/second) in the indoor sequence, and 2.69 milliseconds per frame (37.12 frames/second) in the outdoor sequence on a PC with Pentium 4 CPU of 3.2 GHz and RAM of 1G Bytes. It proves that the proposed algorithm is remarkably fast enough to be applied to real-time surveillance systems even though we used the JM reference software which works relatively slow in comparison with other commercial decoders such as FFMPEG.

Especially, to verify the effect of partial decoding on computational complexity, we compared the computation time in partial decoding with that in full decoding. In the full decoding mode, the frame rate of the proposed algorithm was 20.46 frames/second in the indoor sequence, and 19.17 frames/second in the outdoor sequence. It proves that the partial decoding approach allows us to make the computation time approximately twice faster than that in full decoding.

## 6. CONCLUSION

In this paper, we have presented a novel approach that can rapidly detect and track multiple objects simultaneously from H.264|AVC bitstreams. Our proposed algorithm is considerably practical since it does not merely work well in real time even in general PCs, but also it can efficiently adapt to more natural scenes which include multiple objects, long-time occlusion or the movement of any articulated object. The flexibility is a novelty of the algorithm in that no previous compressed domain algorithms have been verified in such complicated environments.

These features are accomplished by combining the compressed domain approach, which roughly extracts the macroblock-level object region based on probabilistic spatiotemporal macroblock filtering, with the pixel domain approach which precisely refines the object trajectories based on partially decoded color information. In particular, the proposed partial decoding scheme for H.264|AVC bitstreams is proven to be significantly effective in reducing the computational cost. Therefore, this kind of combination can be employed as a fundamental framework to solve the problem of slow processing from which existing vision-based algorithms of pixel domain have suffered. We expect that the proposed algorithm is applicable to many real-time surveillance systems.

In the future works, the proposed algorithm can be extended to better performance. First of all, lots of vision-based techniques can be utilized with the refinement step of our proposed framework to make the tracking system more flexible in a variety of environments. Also, the algorithm could be compatible with video bitstreams which are encoded with the Main profile as well as the Baseline profile of H.264|AVC standard; in other words, B-frames could be handled along with I-frames and P-frames. In addition, it can evolve into the advanced technique which deals with illumination change and silhouette in the compressed domain. The most valuable extension would be to apply the proposed algorithm to a large-scale distributed surveillance system which is designed to process many compressed videos simultaneously and rapidly, and identify and track each object over the camera network.

**REFERENCES**

[1] T. Wiegand, G. J. Sullivan, G. Bjøntegaard, and A. Luthra, "Overview of the H.264|AVC Video Coding Standard," *IEEE Trans. Circuits Syst. Video Technol.*, vol. 13, No. 7, pp. 560–576, July 2003.

[2] A. Aggarwal, S. Biswas, S. Singh, S. Sural, and A.K. Majumdar, "Object Tracking Using Background Subtraction and Motion Estimation in MPEG Videos," *ACCV 2006*, LNCS, vol. 3852, pp. 121-130, Springer, Heidelberg (2006).

[3] S. Ji and H. W. Park, "Moving object segmentation in DCT-based compressed video," *Electronic Letters*, Vol. 36, No. 21, October 2000.

[4] X. -D. Yu, L.-Y. Duan, and Q. Tian, "Robust moving video object segmentation in the mpeg compressed domain," in *Proc. IEEE Int. Conf. Image Processing*, 2003, vol. 3, pp.933-936.

[5] W. Zeng, W. Gao, and D. Zhao, "Automatic moving object extraction in MPEG video," in *Proc. IEEE Int. Symp. Circuits Syst.*, 2003, vol. 2, pp.524-527.

[6] A. Benzougar, P. Bouthemy, and R. Fablet, "MRF-based moving object detection from MPEG coded video," in *Proc. IEEE Int. Conf. Image Processing*, 2001, vol. 3, pp.402-405.

[7] R. Wang, H.-J. Zhung, Y.-Q. Zhang, "A confidence measure based moving object extraction system built for compressed domain," in *Proc. IEEE Int. Symp. Circuits Syst., 2000*, vol. 5, pp.21-24.

[8] O. Sukmarg and K. R. Rao, "Fast object detection and segmentation in MPEG compressed domain," in *Proc. TENCON 2000*, vol. 3, pp.364-368.

[9] H.-L. Eng and K.-K. Ma, "Spatiotemporal segmentation of moving video objects over MPEG compressed domain," in *Proc. IEEE Int. Conf. Multimedia and Expo*, 2000, vol. 3, pp.1531-1534.

[10] M. L. Jamrozik and M. H. Hayes, "A compressed domain video object segmentation system," in *Proc. IEEE Int. Conf. Image Processing*, 2002, vol. 1, pp.113-116.

[11] H. Zen, T. Hasegawa, and S. Ozawa, "Moving object detection from MPEG coded picture," in *Proc. IEEE Int. Conf. Image Processing*, 1999, vol. 4, pp.25-29.

[12] W. Zeng, J. Du, W. Gao, and Q. Huang, "Robust moving object segmentation on H.264|AVC compressed video using the block-based MRF model," *Real-Time Imaging*, vol. 11(4), 2005, pp.290-299.

[13] R. V. Babu, K. R. Ramakrishnan, and S. H. Srinivasan, "Video object segmentation: A compressed domain approach," *IEEE Trans. Circuits Syst. Video Technol.*, vol. 14, No. 4, pp. 462–474, April 2004.

[14] V. Thilak and C. D. Creusere, "Tracking of extended size targets in H.264 compressed video using the probabilistic data association filter," *EUSIPCO 2004*, pp.281-284, September 2004.

[15] S. Treetasanatavorn, U. Rauschenbach, J. Heuer, and A. Kaup, "Bayesian method for motion segmentation and tracking in compressed videos," *DAGM 2005*, LNCS, vol. 3663, pp.277-284, Springer, Heldelberg (2005).

[16] W. You, M.S. H. Sabirin, and M. Kim, "Moving Object Tracking in H.264/AVC bitstream," *MCAM 2007*, LNCS, vol. 4577, pp.483-492, Springer, Heldelberg (2007).

[17] R.C. Kohtaro Ohba, Yoichi Sato and Katsusi Ikeuchi, "Appearance based visual learning and object recognition with illumination invariance," *Third Asian Conference on Computer Vision (ACCV) 1998*, LNCS, vol.1352, pp. 424-431, Springer, Heldelberg (1997).

[18] Fatih Porikli and Huifang Sun, "Compressed domain video object segmentation," Technical Report TR2005-040 of Mitsubishi Electric Research Lab, 2005.

[19] H. Chen, Y. Zhan and F. Qi, "Rapid object tracking on compressed video," in *Proc. 2nd IEEE Pacific Rim Conference on Multimedia*, pp.1066-1071, October 2001.

[20] V. Mezaris, I. Kompatsiaris, E. Kokkinou, and M.G. Strintzis, "Real-time compressed-domain spatiotemporal video segmentation," in *Proc. CBMI03*, pp.373-380, September 2003.

[21] D. Hall, J. Nascimento, P. Ribeiro, E. Andrade, P. Moreno, S. Pesnel, T. List, R. Emonet, R.B. Fisher, J.S. Victor, and J.L. Crowley, "Comparison of target detection algorithms using adaptive background models," in *Proc. 2nd Joint IEEE International Workshop on Visual Surveillance and Performance Evaluation of Tracking and Surveillance*, pp. 113-120, October 2005.